# Deep-learning of Parametric Partial Differential Equations from Sparse and Noisy Data

Hao Xu[1], Dongxiao Zhang[2,*], and Junsheng Zeng[3]

[1] BIC-ESAT, ERE, and SKLTCS, College of Engineering, Peking University, Beijing 100871, P. R. China

[2] School of Environmental Science and Engineering, Southern University of Science and Technology, Shenzhen 518055, P. R. China

[3] Intelligent Energy Lab, Peng Cheng Laboratory, Shenzhen 518000, P. R. China

Data-driven methods have recently made great progress in the discovery of partial differential equations (PDEs) from spatial-temporal data. However, several challenges remain to be solved, including sparse noisy data, incomplete candidate library, and spatially- or temporally-varying coefficients. In this work, a new framework, which combines neural network, genetic algorithm and adaptive methods, is put forward to address all of these challenges simultaneously. In the framework, a trained neural network is utilized to calculate derivatives and generate a large amount of meta-data, which solves the problem of sparse noisy data. Next, genetic algorithm is utilized to discover the form of PDEs and corresponding coefficients with an incomplete candidate library. Finally, a two-step adaptive method is introduced to discover parametric PDEs with spatially- or temporally-varying coefficients. In this method, the structure of a parametric PDE is first discovered, and then the general form of varying coefficients is identified. The proposed algorithm is tested on the Burgers equation, the convection-diffusion equation, the wave equation, and the KdV equation. The results demonstrate that this method is robust to sparse and noisy data, and is able to discover parametric PDEs with an incomplete candidate library.

**Subject areas:** Computational Physics, Machine Learning, PDE Discovery

## I. INTRODUCTION

In recent years, great advancements have been made in data-driven discovery of partial differential equations (PDEs), whose goal is to identify governing equations of underlying physical processes directly from noisy and sparse observation data. Sparse regression methods, including least absolute shrinkage and selection operator (LASSO) [1], sequential threshold ridge regression (STRidge) [2], sparse Bayesian regression [3] and sparse group lasso [4], have been used to obtain parsimonious models from data for various physical systems. However, three major challenges remain to be solved.

The first challenge is sparse noisy data, which results in serious problems in the calculation of spatial and temporal derivatives. Derivative smoothing methods, such as polynomial interpolation, total variation regularized derivative and integral form, have been utilized to reduce the impact of noise [1,5,6]. Meanwhile, neural networks have been introduced to discover PDEs. For example, the physics-informed neural network (PINN) is proposed by Raissi [7] to find corresponding coefficients with high accuracy when the terms of the PDE are known. PDE-NET is developed by Long et al. [8] to discover time-dependent PDEs, although no parsimony in the PDEs is reinforced.

The residual network (ResNet) and the convolutional neural network (CNN) are also utilized to identify PDEs [9–11]. Compared with sparse regression, neural-network-based methods are more flexible and robust to noise due to their ability to calculate derivatives automatically during the back-propagation process. In addition, discovering the weak form of PDEs that is expressed in integral form has attracted increasing attention. Reinbold et al. [12] proposed a framework to identify the weak form of PDEs, and found that the integral form assists to deal with PDEs with high-order derivatives. Gurevich et al. 13] discovered PDEs from noisy data using a combination of sparse regression and weak form. However, noise levels in the abovementioned works are mostly 1% or 5%, which are relatively small. Overall, when faced with high noise levels (e.g., 10% or more) or very sparse data, the accuracy of these methods still needs to be improved.

The second challenge is incomplete candidate library. All methods mentioned above are constructed on the basis of a complete candidate library, which means that true PDE terms are contained in the candidate library. However, it is difficult to construct a complete candidate library in cases in which little prior knowledge is known. In addition, as the size of the candidate library grows, the difficulty in obtaining sparsity also increases, which may lead to discovering an incorrect PDE. To overcome this challenge, genetic algorithm is introduced to discover PDEs with an incomplete candidate library. For example, an evolutionary optimization-based method is proposed by Maslyaev et al. [14] to discover PDEs with an incomplete candidate library, and achieves better performance than STRidge. Meanwhile, a genetic-algorithm-based method is put forward by Atkinson et al. [15] to identify free-form PDEs. Although these methods possess the advantage of discovering PDEs with an incomplete candidate library, they are unable to handle sparse noisy data because derivatives are calculated by finite difference, which is not robust to noise and requires that data are distributed on a regular grid.

The final challenge is spatially- or temporally-varying coefficients. In some cases, coefficients of the PDE are variable instead of constant, which poses a challenge to PDE discovery. All of the abovementioned methods aim to address the problem of PDEs with constant parameters, but they are unable to handle varying coefficients. Some recent works attempt to discover PDEs with varying parameters. Rudy et al. [16] proposed a new framework, called sequential grouped threshold ridge regression (SGTR), to discover parametric PDEs. In their work, a separate regression is constructed for each time step to obtain corresponding time-varying coefficients. However, this method only obtains coefficients corresponding to each single point, but cannot obtain a general form of varying coefficients. Moreover, SGTR is constructed on the basis of a complete candidate library and is only robust to a low level of noise. Xiong et al. [9] changed the parameter of inhomogeneous terms from a single number to a length-adaptive tensor, and obtained inhomogeneous coefficients by the gradient descent method of a trained neural network. This method, however, is also sensitive to noise due to the error produced in the process of training the inhomogeneous parameter tensor in the neural network. Furthermore, although the discovered varying coefficients have an explicit form, they are expressed as a kind of trigonometric series, so that the true form of coefficients may not be discovered. All of these methods are also based on the assumption that the type of varying parameter is known. In other words, prior knowledge about whether the coefficients are spatially-varying or temporally-varying is required, which is unrealistic in practice.

To address the first challenge, our earlier work [17] proposed a framework, called DL-PDE, which combines neural network and sparse regression. Specifically, it utilizes a trained neural network to generate a large amount of meta-data and calculate the corresponding derivatives, and then employs sparse regression, such as STRidge, to obtain the PDE form and corresponding coefficients. Numerical experiments have demonstrated that this method performs well when data are noisy and sparse. To overcome the second challenge, a new framework combining neural network and genetic algorithm, called DLGA-PDE, is proposed on the basis of DL-PDE [18]. Through mutation and cross-over in the genetic algorithm, DLGA-PDE can generate infinite combinations of basic genes, and performs well with an incomplete candidate library. In this work, a novel framework combining adaptive methods and DLGA-PDE, called Adaptive DLGA-PDE, is developed to simultaneously handle all three challenges mentioned above. In this method, a two-step DLGA-PDE is performed to discover parametric PDEs from sparse noisy data with an incomplete library. It utilizes DLGA-PDE (Structure) to identify the type of varying parameters and discover the structure of PDE from local meta-data within different windows, and utilizes DLGA-PDE (Coefficients) to identify the general form of varying coefficients from global meta-data. The proposed Adaptive DLGA-PDE is tested with the Burgers equation, the convection-diffusion equation, the wave equation, and the KdV equation for proof-of-concept and sensitivity studies, and satisfactory results are obtained.

The remainder of this paper is organized as follows. In Section II, problem settings and the procedure of the proposed algorithm are presented. In Section III, the proposed algorithm is applied to discover four parametric PDEs. The performance of the Adaptive DLGA-PDE is tested in the case of sparse noisy data and incomplete candidate library, respectively. Finally, Section IV summarizes the advantages and disadvantages of the proposed algorithm, and suggests directions for future research.

## II METHOD

### A. Parametric PDE discovery

In this work, parametric PDEs with spatially- or temporally-varying coefficients are considered. The PDE with spatially-varying coefficients is taken as an example here, and the PDE with temporally-varying coefficients can be expressed similarly. The form of the parametric PDE can be written as follows:

$$u_T = N(\Phi(u);[\alpha_1(x),\alpha_2(x),...,\alpha_n(x)]), \qquad (1)$$

with

$$\Phi(u) = [u, u^2, u_x, u_{xx},...,uu_x, uu_{xx},...,], \qquad (2)$$

where $u_T$ refers to different orders of derivatives of $u$ with respect to $t$, such as $u_t$ and $u_{tt}$; $\Phi(u)$ denotes the candidate library of potential terms; $\alpha_i(x)$ ($i$=1,2,...,$n$) denotes the vector of varying coefficients, with $n$ being the size of the candidate library; and $N(\cdot)$ denotes the linear combination of terms in $\Phi(u)$ and corresponding vector of varying coefficients $\alpha_i(x)$.

For spatial-temporal observation data or meta-data, denoted as $u(x_i,t_j), i=1,2,...,N_x, j=1,2,...,N_t$, Eq. (1) can be rewritten as:

$$u_T(x_i,t_j) = \begin{bmatrix} u_x(x_i,t_j) & \cdots & uu_x(x_i,t_j) & \cdots \end{bmatrix} \begin{bmatrix} \alpha_1(x_i) \\ \alpha_2(x_i) \\ \vdots \\ \alpha_n(x_i) \end{bmatrix}, \ i=1,2,...N_x, \ j=1,2,...N_t, \quad (3)$$

where $N_x$ is the number of $x$; $N_t$ is the number of $t$; and $n$ is the number of possible terms. Eq. (3) can be expressed as follows:

$$U_T = \Theta(u) \cdot \Lambda(x). \quad (4)$$

The discovery of parametric PDE aims to identify nonzero PDE terms from $\Theta(u)$ and the general form of varying coefficients $\Lambda(x)$. Here, a parsimonious model is expected to be obtained since, for most partial differential equations, only a small number of terms exist, and thus with nonzero coefficients, in Eq. (1) and (2).

### B. Architecture of Adaptive DLGA-PDE

In this work, a novel framework, called Adaptive DLGA-PDE, is proposed to discover parametric PDEs from sparse noisy data with an incomplete candidate library. This algorithm constitutes a combination of adaptive methods and DLGA-PDE. Due to the difficulty in discovering the PDE form and spatially- or temporally-varying coefficients at the same time, a two-step adaptive method is utilized. The structure of the PDE is discovered firstly, and then the general form of varying coefficients is identified.

In this framework, sparse noisy data are used to train a fully-connected feedback neural network. The trained neural network is then utilized to calculate derivatives and generate meta-data. Here, two types of meta-data, including local meta-data and global meta-data, are generated. Local meta-data are generated within a local window, which is a small part of the entire domain, and global meta-data are generated within nearly the whole domain. For local meta-data, DLGA-PDE is performed to discover the structure of PDEs, although the coefficients may be either incorrect or non-representative, because no global constraint is reinforced at this step. This process is called DLGA-PDE (Structure). Then, the type of varying parameter (spatially-varying or temporally-varying) is identified by comparing the spatial and temporal stability of DLGA-PDE (Structure). Subsequently, corresponding coefficients at each point ($x$ or $t$) in the global domain are calculated based on the discovered structure of PDE by global optimization of the neural network with global meta-data. Finally, a new DLGA-PDE method that is different from DLGA-PDE (Structure) is used to discover the general form of the calculated varying coefficients. This process is called DLGA-PDE (Coefficients). The architecture of Adaptive DLGA-PDE is shown in Fig. 1.

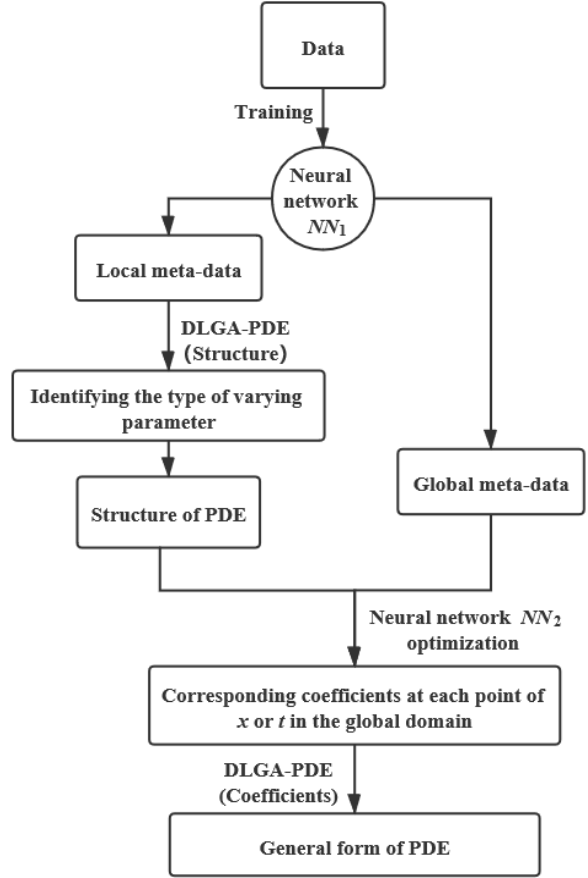

FIG. 1. Workflow scheme of Adaptive DLGA-PDE.

## C. Neural network

In this work, a deep fully-connected artificial neural network $NN_1(x,t;\theta)$ is trained to approximate $u(x,t)$. The structure of a typical artificial neural network is shown in Fig. 2. Input of the neural network is spatial-temporal observation data, and output is $NN_1(x_i,t_j;\theta)$, where $\theta$ refers to the parameters of the neural network, including weights and bias.

The neural network $NN_1(x,t;\theta)$ is trained by minimizing the loss function:

$$Loss(\theta) = \sum_{j=1}^{N_t}\sum_{i=1}^{N_x}[u(x_i,t_j) - NN_1(x_i,t_j;\theta)]^2, \tag{5}$$

where $N_x$ is the number of $x$; and $N_t$ is the number of $t$. The early termination method is utilized to prevent over-fitting. When the training process has been completed, derivatives can be calculated via automatic differentiation. Meanwhile, the trained neural network $NN_1(x,t;\theta)$ serves as a surrogate model for the underlying physical system that can be employed to generate meta-data. Here, the neural network is briefly introduced, additional details of which can be found in Xu et al. [17].

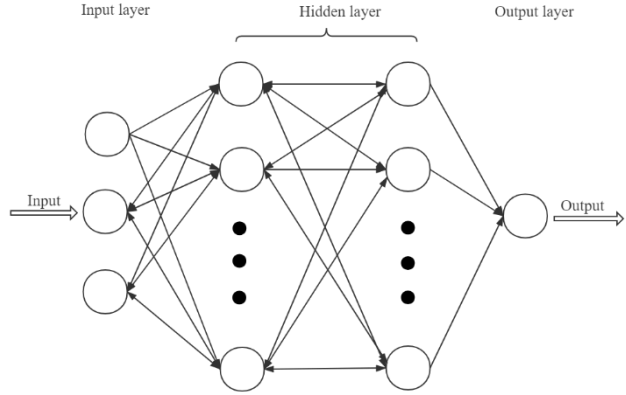

**FIG. 2.** Structure of a typical artificial neural network.

### D. Genetic algorithm

The genetic algorithm is an important part of Adaptive DLGA-PDE, whose function is discovering the form of PDEs with an incomplete candidate library. With a unique digitization method, terms in the PDE can be expressed by genes. Through mutation and cross-over, the genetic algorithm can produce infinite combinations of genes, which greatly expands the search scope of possible PDE terms. In this part, we will briefly introduce the procedure of genetic algorithm utilized in Adaptive DLGA-PDE, including translation, cross-over, mutation, fitness calculation, and selection. Specially, a new principle, called winner-take-all, is introduced to prevent being constrained in a local minimum. Additional details about the genetic algorithm can be found in Xu et al. [18].

#### 1. Translation

To digitize PDEs, a principle of translation from the structure of PDEs to genomes is proposed. Firstly, numbers are used to denote different orders of derivatives. For example, 0 refers to $u$, 1 refers to $u_x$ or $u_t$, and 2 refers to $u_{xx}$ or $u_{tt}$. It is defined as the gene, which is the smallest unit in the genetic algorithm. The combination of genes is defined to be a gene module. It is assumed that there is only multiplication in a module. For example, [0,1] refers to $uu_x$, and [1,1,2] refers to $u_x^2 u_{xx}$. The combination of gene modules is defined as the genome. A genome has two parts, which represent left-hand terms and right-hand terms in the PDEs, respectively. It is also assumed that the left-hand terms are derivatives with respect to $t$, and the right-hand terms are derivatives with respect to $x$. Gene modules are connected by addition. For example, [1],{[0,1],[2]} refers to the structure of PDE $u_t = uu_x + u_{xx}$. Here, for the sake of distinction, right-hand terms are placed in the brace. With the special principle of translation, we can generate a large number of genomes from several basic genes. Each genome corresponds to a specific PDE.

#### 2. Cross-over

Cross-over is the process in which two parents exchange certain gene modules to produce their children. It is an important way for parents to transfer their genes to the next generation. In this work, the probability of cross-over is 80%.

### 3. Mutation

Mutation is a key process for producing new genes, which is critical in the genetic algorithm. There are many ways of mutation. Here, three main ways, including order mutation, add-gene mutation and delete-gene mutation, are introduced. In order mutation, certain genes in the genomes will be changed by reducing 1. Particularly, 0 can be changed to the largest number. For example, genome [1],{[0],[1,3]} may be transformed to be [1],{[3],[1,2]} if 3 is the largest number. In add-gene mutation, a new random gene module is added into the genome. In delete-gene mutation, a certain gene module is deleted from the genome. These three ways of mutation constantly generate new genomes during the evolution process, thereby avoiding the local minimum. In this work, the chance of mutation is set to be 80%.

### 4. Fitness calculation

Fitness refers to the viability of a genome in the environment, measured as the quality of the genome. Fitness is usually calculated by a fitness function, which is defined in this work as:

$$Fitness = MSE_1 + \varepsilon \cdot len(genome), \qquad (6)$$

$$MSE_1 = \frac{\sum_{j=1}^{N_t}\sum_{i=1}^{N_x}\left|U_L(x_i,t_j) - \vec{\beta_1} U_R(x_i,t_j)\right|^2}{N_x N_t},$$

where $U_L(x_i,t_j)$ is the value of the left-hand term, which is a $1\times 1$ vector; $U_R(x_i,t_j)$ is the value of right-hand terms in the proposed PDE, which is a $n\times 1$ vector; n is the number of PDE terms; and $N_x$ and $N_t$ are the number of $x$ and $t$, respectively. For each genome and its corresponding structure of equation, the vector of coefficients $\vec{\beta_1}$ with the size of $1\times n$ for the right-hand terms is obtained by the least square regression, and the mean squared error $MSE_1$ is then calculated. To prevent over-fitting, the $l_0$ penalty is utilized. Here, *len(genome)* is the length of the genome, and $\varepsilon$ is a hyper-parameter, which is chosen according to the magnitude of $MSE_1$. It is worth mentioning that the genetic algorithm uses a derivative-independent way to optimize fitness, and thus the $l_0$ penalty, which is simpler and more effective, can be applied here. In this work, the smaller is the fitness function, the better is the genome.

### 5. Selection

In this work, each parent genome cross-overs twice, and the best half of children in each generation is selected to be the next generation of parents. After several generations, the best child whose fitness function is the smallest is the best model.

### 6. Principle of winner-take-all

In some cases, the genetic algorithm may easily fall into a local minimum, which is detrimental to finding the best equation. Repeated trial is an option, but is tedious and unstable. Consequently, in this work, a principle called winner-take-all is proposed. This means that the winner will take all of the benefits, while the others receive none. In the winner-take-all principle, except for the best genome, all other genomes will be replaced by new random genomes under a certain probability.

This principle will assist to avoid the local minimum and accelerate the rate of convergence. In this work, the principle of winner-take-all is utilized in DLGA-PDE (Coefficients).

### E. Procedure of DLGA-PDE (Structure) and DLGA-PDE (Coefficients)

In this subsection, the procedure of DLGA-PDE (Structure) and DLGA-PDE (Coefficients) are introduced. DLGA-PDE (Structure) aims to discover the structure of the PDE from local meta-data within a window, while DLGA-PDE (Coefficients) aims to identify the general form of spatially- or temporally-varying coefficients. The genetic algorithm is utilized in both processes, but the settings are dissimilar.

#### 1. DLGA-PDE (Structure)

In DLGA-PDE (Structure), basic genes of DLGA-PDE are $[u_t(1),u_{tt}(2)],\{u(0),u_x(1),u_{xx}(2),u_{xxx}(3)\}$. 100 genomes are produced from basic genes. For each genome, it cross-overs twice and 200 children are generated. For each child, mutation takes place under the probability of 20%. It is worth noting that three ways of mutation take place independently. After mutation, the fitness of each child is calculated according to the fitness function and local meta-data. Children are then sorted according to their fitness, and the first half of the best children are left as the parents of the next generation. This process will repeat until convergence or a maximum number of generations is reached. The best child in the last generation is the discovered structure. The work-flow of DLGA-PDE (Structure) is shown in Fig. 3.

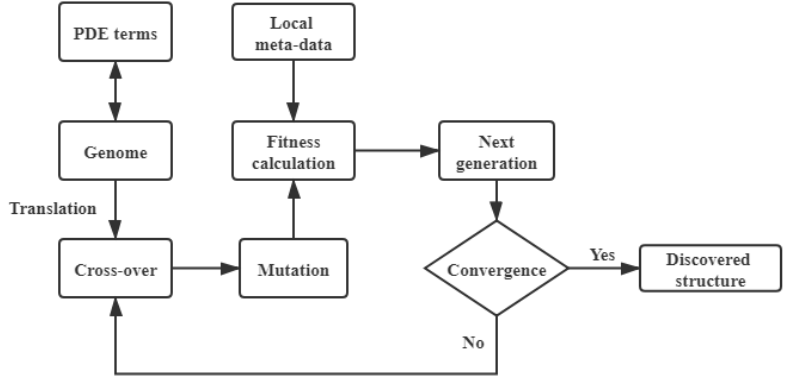

**FIG. 3.** Work-flow of DLGA-PDE (Structure).

#### 2. Identifying the type of varying parameters and discovering the best structure

Although the structure of PDE can be discovered by DLGA-PDE (Structure) for a local window in the above, it remains unclear whether the coefficients vary spatially or temporally. Therefore, the type of varying coefficients has to be identified.

Firstly, the coefficients are assumed to be spatially-varying. As shown in Fig. 4, a large number of local meta-data is generated from the neural network $NN_1(x,t;\theta)$ within several local windows. For each local window, DLGA-PDE (Structure) is performed, and a respective structure is discovered. Among these structures, the structure which occurs stably and most frequently is termed the best (or most possible) structure. Then, we may examine the spatial stability of DLGA-PDE (Structure), which is defined as follows:

$$S_x = \frac{N_x^{best}}{N_x} \qquad (7)$$

where $N_x$ is the number of local windows; and $N_x^{best}$ is the number of occurrences of the best structure in space. The coefficients are then assumed to be temporally-varying. The respective best structure in time and the corresponding temporal stability $S_t$ can be discovered in a similar manner.

If the assumption of either spatially- or temporally-varying coefficients is correct, the type of varying coefficients does not change substantially in the local windows, and the structures discovered in multiple local windows will be more stable. In contrast, if the assumption is incorrect, DLGA-PDE (Structure) will be relatively unstable because it will lead to different structures of PDE within different local windows. Therefore, if $S_x > S_t$, the coefficient is spatially-varying; otherwise, the coefficient is temporally-varying. After identifying the type of varying parameter, the finally discovered structure is the corresponding best structure.

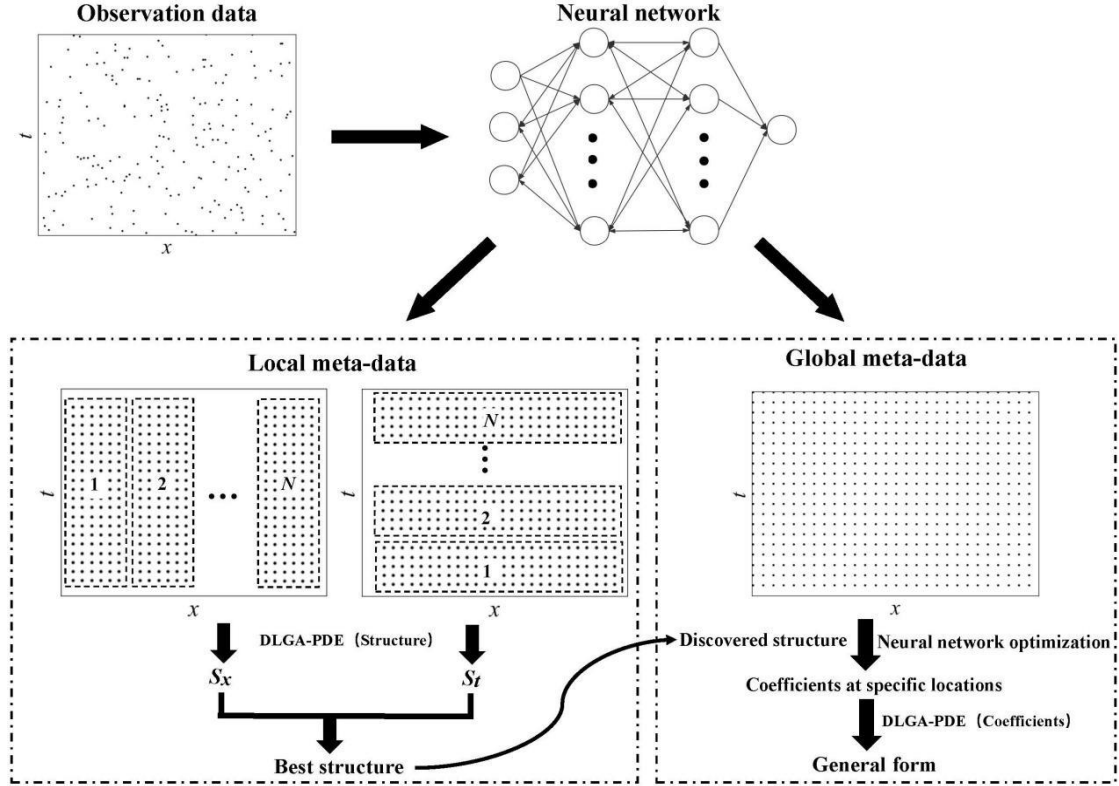

**FIG. 4.** Diagram of the proposed algorithm, with black dots referring to data. The observation data are sparse and noisy, while meta-data produced by the neural network are dense.

*3. Calculating varying coefficients*

Although the structure of the PDE has been discovered, the particular values and the corresponding expression of varying coefficients are unknown. As shown in Fig. 4, a large number of global meta-data are generated within nearly the entire domain. With the discovered structure and global meta-data, the process of calculating corresponding coefficients at each point of $x$ or $t$ in the global domain constitutes an inverse modelling problem, which can be solved by neural network methods [7] or data-assimilation methods, such as ensemble Kalman filter (EnKF) [19,20]. In this work, corresponding values of the varying coefficients are calculated by global optimization of the neural network. The neural network $NN_2(x,t;\theta,\Lambda)$ is constructed in a similar way to $NN_1(x,t;\theta)$, while the loss function is different. Here, the loss function is defined as follows:

$$Loss(\Lambda) = \sum_{j=1}^{N_t}\sum_{i=1}^{N_x}\left[U_L^{global}(x_i^{global},t_j^{global}) - \Lambda U_R^{global}(x_i^{global},t_j^{global}))\right]^2, \qquad (8)$$

with:

$$\Lambda = [\alpha_1, \alpha_2, ... \alpha_n]; \quad \alpha_i = \begin{bmatrix} \alpha_i(x_1) \\ \alpha_i(x_2) \\ \vdots \\ \alpha_i(x_{N_x}) \end{bmatrix}, \quad i=1,2,...,n, \qquad (9)$$

where $x_i^{global}$ and $t_j^{global}$ are the points of $x$ and $t$ in the global meta-data, respectively; $N_x$ and $N_t$ are the number of $x$ and $t$ in the global meta-data, respectively; $n$ is the number of terms in the discovered structure; $U_L^{global}$ and $U_R^{global}$ are the left-hand terms and right-hand terms of the discovered structure, respectively; and $\Lambda$ is the parameter tensor, with $\alpha_i (i=1,2,...,n)$ being the corresponding varying coefficient at each $x$ or $t$ in the global domain for each term. Since derivatives in these terms can be calculated from automatic differentiation of the previous trained neural network $NN_1(x,t;\theta)$, the values of $U_L^{global}$ and $U_L^{global}$ are known. This means that the only unknown in the loss function is the parameter tensor $\Lambda$, which can be optimized via the neural network $NN_2(x,t;\theta,\Lambda)$.

The input of the neural network is $x$ and $t$ of the global meta-data. Initially, all elements in the tensor $\alpha_i (i=1,2,...,n)$ are set to be 1. During the training process, the parameter $\Lambda$ and the network parameters $\theta$, including weight and bias, are optimized to minimize the loss function at the same time. After the training process, the loss function has been minimized, and the corresponding varying coefficient $\alpha_i (i=1,2,...,n)$ can be obtained via optimization. It is worth noting that, although the corresponding coefficients at specific locations are obtained in this step, the general form, which is more representative, remains undetermined.

*4. DLGA-PDE (Coefficients)*

In DLGA-PDE (Coefficients), it is assumed that the general form of varying coefficients is comprised of elementary functions, such as $k_1x^n$, $\sin(k_2x)$, $\cos(k_3x)$ and $e^{k_4x}$, and $n$ is a positive integer and $k_i$ ($i$=1,2,3,4) is a real number. As a consequence, basic genes of DLGA-PDE (Coefficients) are $[\alpha_i]\{1(0),k_1x(1),\sin(k_2x)(2),\cos(k_3x)(3),e^{k_4x}(4)\}$ for spatial-varying coefficients and $[\alpha_i]\{1(0),k_1t(1),\sin(k_2t)(2),\cos(k_3t)(3),e^{k_4t}(4)\}$ for temporal-varying coefficients. Here, $\alpha_i$ refers to the corresponding calculated coefficients by global optimization of the neural network in the above.

With the definition of basic genes, genomes can be generated. Different from DLGA-PDE

(Structure), genomes in DLGA-PDE (Coefficients) have two genome sequences with the same size. For example:

$$Genomes: \begin{matrix} \{[t,\sin(t)],[\cos(t)]\} \\ \{[k_1,k_2],[k_3]\} \end{matrix} \longleftrightarrow Functions: k_1 t \sin(k_2 t) + \cos(k_3 t)$$

Here, the upper sequence refers to possible functions in the general form of varying parameters, and the lower one refers to corresponding coefficients of these terms, respectively. In this work, $k_i = delta \ast N$ ($i$=1,2,3,...,), where $delta$ is the interval of $k_i$ that is set to be 0.001, and $N$ is a random integer from [-10000,10000]. Therefore, the value range of $k_i$ is [-10,10], which is sufficient for most common parametric PDEs. For more complex situations, in which coefficients may vary with a higher frequency, a larger value range of $k_i$ can be adopted.

After generating genomes, the principle of winner-take-all is performed prior to cross-over and mutation. In this step, the best genome remains, while other genomes are replaced by new random genomes under a certain probability, which is 80% in this work. Due to multiple possibilities produced by these two genome sequences, the genetic algorithm is very easily constrained to a local minimum, which may lead to incorrect forms of the equation. Indeed, the principle of winner-take-all can effectively avoid the local minimum without repeated trails.

Cross-over and mutation are then performed. It is worth noting that the two genome sequences cross-over and mutate synchronously. Particularly, $k_i$ mutates via randomly selecting a new value to replace itself. Subsequently, fitness is calculated. For each genome, the corresponding function can be obtained by translation, and terms in the function can be calculated. Here, the fitness function is slightly different, which is defined as:

$$Fitness = MSE_2 + \varepsilon_1 \cdot len(genome) + \varepsilon_2 \cdot if\_not\_constant, \qquad (10)$$

$$MSE_2 = \frac{\sum_{j=1}^{N} \left| \alpha_i(x_j) - \vec{\beta}_2 F(x_j) \right|^2}{N},$$

where $F(x_j)$ is the vector of possible functions, with the size $n \times 1$; $n$ is the number of terms in the function; and $N$ is the amount of $x$ in global meta-data. The vector $\vec{\beta}_2$ of size $1 \times n$ is coefficients of the corresponding functions that are obtained by the least square regression, and the mean squared error $MSE_2$ is then calculated. Here, $if\_not\_constant$ is the index for distinction between constant and varying coefficients. If the discovered form of the coefficient is a constant, $if\_not\_constant$=0; otherwise, $if\_not\_constant$=1. This penalty is applied to avoid overfitting constant coefficients by a series of complex formulas. $\varepsilon_1$ and $\varepsilon_2$ are hyper-parameters, and are chosen according to the magnitude of $MSE_2$. Here, the spatially-varying coefficient is taken as an example, and the fitness for the temporally-varying coefficient can be calculated similarly.

Finally, selection is done according to fitness. When the evolution has converged, the best child is the discovered general form of varying coefficients. The work-flow of DLGA-PDE (Coefficients) is presented in Fig. 5.

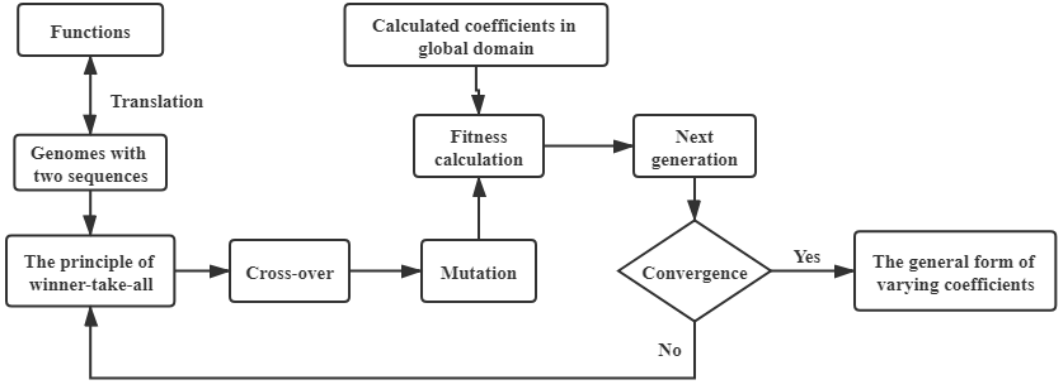

**FIG. 5.** Workflow of DLGA-PDE (Coefficients).

## III. RESULTS

To test the performance of Adaptive DLGA-PDE, three PDEs with spatially- or temporally-varying coefficients are considered, including the Burgers equation, the convection-diffusion equation, and the wave equation. Meanwhile, the KdV equation, which has high-order derivatives, is also investigated. Although the general form of varying coefficients in the KdV equation can be discovered, accuracy needs to be improved. Additional details and discussions are provided in Appendix A.

In this work, it is supposed that the type of coefficients is consistent in the PDE, which means the PDE only has temporally-varying coefficients or spatially-varying coefficients. Two neural networks $NN_1(x,t;\theta)$ and $NN_2(x,t;\theta,\Lambda)$ both have five layers, an input layer, an output layer and three hidden layers, with 50 neurons in each hidden layer. The activation function is $\sin(x)$. The population size of genomes in DLGA-PDE (Structure) is 200, and the number of maximum generations is 100. Moreover, the population size of genomes in DLGA-PDE (Coefficients) is 1000, and the number of maximum generations is 200. From the sensitivity analysis, which is detailed in Appendix B1, for a wide range of local window size, the performance of DLGA-PDE (Structure) is stable. Therefore, in this work, the local window size is fixed to be 10% of the entire domain.

### A. Discovery of parametric PDEs with noisy data

#### 1. Parametric Burgers equation

Parametric Burgers equation with a temporally-varying coefficient $a(t)$ is first considered, whose form is expressed as follows:

$$u_t = -uu_x + a(t)u_{xx}, \qquad (11)$$

$$a(t) = 0.02e^t,$$

To generate training data, the Burgers equation is solved numerically using the finite difference method with the initial condition $(1.5-x)\sin(2\pi x)$ and boundary condition $u(0,t)=u(1,t)=0$, $t>0$. There are 200 temporal observation steps in the domain $t \in [0,1)$ and 251 spatial observation steps in the domain $x \in [0,1]$, and thus the total number of data points is 50,200, of which 25,000 data are randomly chosen to train the neural network $NN_1(x,t;\theta)$. In addition to the basic case with clean data, four noise levels, including 1%, 5%, 10% and 15%, are added to the data.

To identify the type of varying parameters and discover the structure of PDE, local meta-data are generated from multiple different local windows. With the assumption of temporal-varying coefficients, the local windows are $t \in [0.1 \cdot n, 0.1 \cdot (n+1)]$, ($n$=0,1,...,9), for each of which there are 400 temporal observation steps and 400 spatial observation steps in the domain $x \in [0.1, 0.9]$; with the assumption of spatially-varying coefficients, the local windows are $x \in [0.1 \cdot n, 0.1 \cdot (n+1)]$, ($n$=0,1,...,9), for each of which there are 400 spatial observation steps and 400 temporal observation steps in the domain $t \in [0.1, 0.9]$. In total, there are 160,000 local meta-data for each local window. DLGA-PDE (Structure) is performed on each local window to calculate $S_x$ and $S_t$. Sensitivity analysis shows that DLGA-PDE (Structure) is insensitive to local window position with the correct assumption, but is sensitive to the local window position with the incorrect assumption, which is detailed in Appendix B2. It can be found that $S_t > S_x$ in all experiments, which means that the coefficients are temporally-varying. Therefore, the best structure is the structure which occurs most frequently with the assumption of temporally-varying coefficients, which is shown in Table I. It is obvious that the discovered best structure is correct.

For global meta-data, there are 300 temporal observation steps in the domain $t \in [0.2, 0.9]$ and 400 spatial observation steps in the domain $x \in [0.1, 0.9]$. In total, 120,000 global meta-data are generated. The values of varying coefficients are calculated by global optimization of the neural network, and DLGA-PDE (Coefficients) is performed to identify the general form of the varying coefficients. The results are shown in Table I. It is found that the general form of varying coefficients is discovered with high accuracy, even if the noise level is 15%. The calculated varying coefficients and the identified general form are shown in Fig. 6(a) and (b) to better demonstrate the effect of DLGA-PDE (Coefficients). From the figure, it can be seen that the varying coefficients calculated by global optimization are robust to noise and relatively accurate. Furthermore, the curve that represents the identified general form is smoother and more accurate.

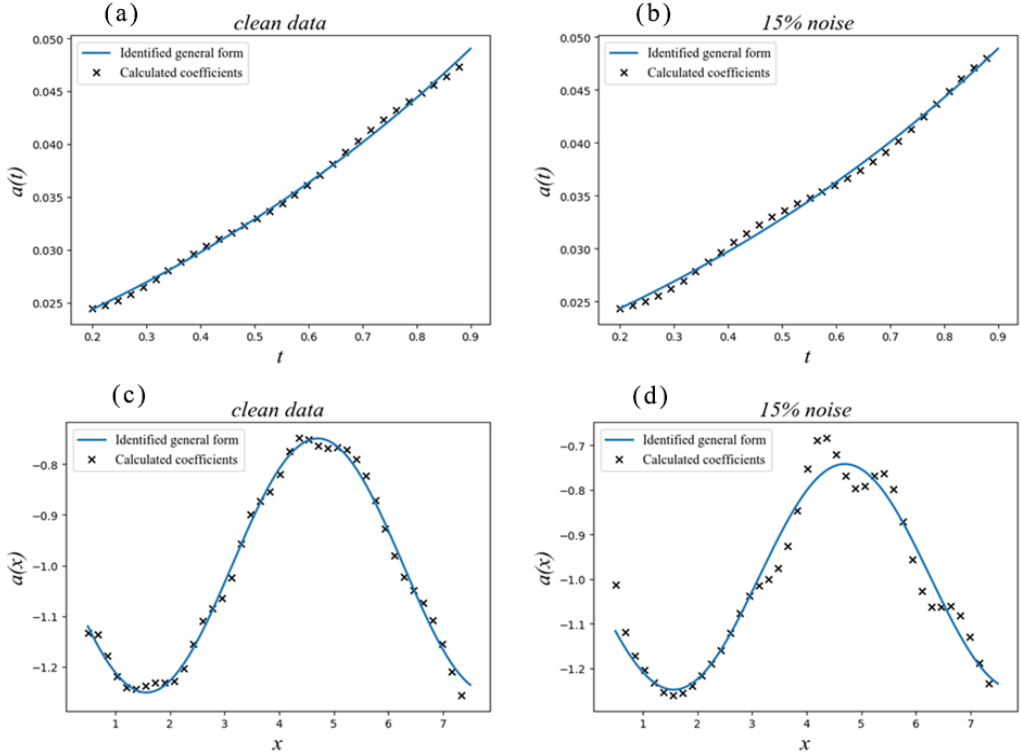

**FIG. 6**. Scatterplot of calculated coefficients, and the curve of the identified general form for the Burgers equation and the convection-diffusion equation with different levels of noise. (a) and (b) is clean data and 15% noise (Burgers equation), respectively; (c) and (d) is clean data and 15% noise (convection-diffusion equation), respectively.

To illustrate the accuracy of discovered PDEs, error is defined as:

$$error = \frac{\sum_{j=1}^{N_t}\sum_{i=1}^{N_x}|u(x_i,t_j)-u'(x_i,t_j)|^2}{N_x N_t} \quad (12)$$

where $u(x_i,t_j)$ is the solution of correct PDE; $u'(x_i,t_j)$ is the solution of discovered PDE; $N_x$ and $N_t$ are the number of $x$ and $t$, respectively. It is obvious that the error is relatively small, which demonstrates that Adaptive DLGA-PDE has discovered the parametric Burgers equation with high accuracy.

**TABLE I.** Structure of parametric Burgers equation found using DLGA-PDE (Structure), and general parametric form found using DLGA-PDE (Coefficients), with clean and noisy data.

| Noise Level | $S_x$ | $S_t$ | Discovered Structure | Learned General Form by DLGA-PDE (Coefficients) | Error |
|---|---|---|---|---|---|
| Correct PDE | | | | $u_t = -uu_x + 0.02e^t u_{xx}$ | |
| Clean Data | 0.3 | 0.8 | $u_t, uu_x, u_{xx}$ | $u_t = -0.996uu_x + 0.01995e^{0.999t}u_{xx}$ | $1.15\times 10^{-6}$ |
| 1% Noise | 0.2 | 0.7 | $u_t, uu_x, u_{xx}$ | $u_t = -0.995uu_x + 0.01995e^{0.999t}u_{xx}$ | $1.69\times 10^{-6}$ |
| 5% Noise | 0.2 | 0.7 | $u_t, uu_x, u_{xx}$ | $u_t = -0.999uu_x + 0.02004e^{0.994t}u_{xx}$ | $9.06\times 10^{-8}$ |
| 10% Noise | 0.2 | 0.8 | $u_t, uu_x, u_{xx}$ | $u_t = -1.001uu_x + 0.01994e^{1.001t}u_{xx}$ | $1.48\times 10^{-7}$ |
| 15% Noise | 0.2 | 0.7 | $u_t, uu_x, u_{xx}$ | $u_t = -0.994uu_x + 0.01995e^{0.996t}u_{xx}$ | $2.57\times 10^{-6}$ |

*2. Parametric convection-diffusion equation*

Next, the parametric convection-diffusion equation with a spatially-varying coefficient $a(x)$ is considered, whose form is expressed as:

$$u_t = a(x)u_x + u_{xx}, \quad (13)$$
$$a(x) = -(1+0.25\sin(x)),$$

To generate training data, the convection-diffusion equation is solved numerically using the precise integration method with the initial condition (8-$x$)sin($x$) and boundary condition $u(0,t)=u(8,t)=0$, $t>0$. There are 250 temporal observation steps in $t \in [0,5)$ and 201 spatial observation steps in $x \in [0,8]$, and thus the total number of data points is 50,250, of which 25,000 data are randomly chosen to train the neural network. In addition to the basic case with clean data, four noise levels, including 1%, 5%, 10% and 15%, are added to the data.

For local meta-data, with the assumption of spatially-varying coefficients, local windows are $t \in [0.5 \cdot n, 0.5 \cdot (n+1)]$, (n=0,1,...,9), for each of which there are 400 temporal observation steps and 400 spatial observation steps in the domain $x \in [0.8, 7.2]$; with the assumption of spatially-varying coefficients, local windows are $x \in [0.8 \cdot n, 0.8 \cdot (n+1)]$, (n=0,1,...,9), for each of which there are 400 spatial observation steps and 400 temporal observation steps in the domain $t \in [0.5, 4.5]$. $S_x$ and $S_t$ calculated are given in Table II, and it can be found that $S_x > S_t$ in all experiments, which means that the coefficients are spatially-varying. Meanwhile, the best structure is obtained, which is shown in Table II.

For global meta-data, there are 300 temporal observation steps in the domain $t \in [0.2, 4.8]$ and 400 spatial observation steps in the domain $x \in [0.5, 7.5]$. The general form is identified, and the results are presented in Table II. The calculated varying coefficients and the identified general form are shown in Fig. 6(c) and (d). It can be seen that, although the calculated coefficients have some error when the noise level is 15%, the general form can be discovered by DLGA-PDE (Coefficients) accurately.

**TABLE II.** Structure of parametric convection-diffusion equation found using DLGA-PDE (Structure), and general parametric form found using DLGA-PDE (Coefficients), with clean and noisy data.

| Noise Level | $S_x$ | $S_t$ | Discovered Structure | Learned General Form by DLGA-PDE (Coefficients) | Error |
|---|---|---|---|---|---|
| Correct PDE | | | | $u_t = -(1+0.25\sin(x))u_x + u_{xx}$ | |
| Clean Data | 0.9 | 0.3 | $u_t, u_x, u_{xx}$ | $u_t = -(1.000+0.251\sin(1.001x))u_x + 0.999u_{xx}$ | $1.12 \times 10^{-6}$ |
| 1% Noise | 0.9 | 0.2 | $u_t, u_x, u_{xx}$ | $u_t = -(1.001+0.250\sin(1.003x))u_x + 1.003u_{xx}$ | $5.39 \times 10^{-6}$ |
| 5% Noise | 0.9 | 0.3 | $u_t, u_x, u_{xx}$ | $u_t = -(0.998+0.247\sin(1.005x))u_x + 1.000u_{xx}$ | $7.06 \times 10^{-6}$ |
| 10% Noise | 0.9 | 0.2 | $u_t, u_x, u_{xx}$ | $u_t = -(0.997+0.246\sin(1.007x))u_x + 0.993u_{xx}$ | $8.36 \times 10^{-6}$ |
| 15% Noise | 0.9 | 0.3 | $u_t, u_x, u_{xx}$ | $u_t = -(0.995+0.253\sin(1.003x))u_x + 0.989u_{xx}$ | $2.12 \times 10^{-5}$ |

*3. Parametric wave equation*

Finally, the parametric wave equation with a temporally-varying coefficient $A(t)$ is considered, whose form is expressed as follows:

$$u_{tt} = A(t)u_{xx}, \quad (14)$$

$$A(t) = 1 + 0.25\cos(1.5t),$$

To generate training data, the wave equation is solved numerically using the precise integration method, with the initial condition

$$u(x,0) = \begin{cases} \dfrac{\sin(2x)}{2}, & 0 \leq x < \dfrac{\pi}{2} \\ 0, & \dfrac{\pi}{2} \leq x \leq \pi \end{cases} ; \quad \dfrac{\partial u}{\partial t}(x,0) = 0$$

and boundary condition $u(0,t)=u(3,t)=0$, $t>0$. There are 400 temporal observation steps in the domain $t \in [0,6)$ and 161 spatial observation steps in the domain $x \in [0,3]$, and thus the total number of data points is 64,400, of which 30,000 data are randomly chosen to train the neural network. In addition to the basic case with clean data, four noise levels, including 1%, 5%, 10% and 15%, are added to the data.

For local meta-data, with the assumption of temporal-varying coefficients, the local windows are $t \in [0.6 \cdot n, 0.6 \cdot (n+1)]$, ($n$=0,1,...,9), for each of which there are 400 temporal observation steps and 400 spatial observation steps in the domain $x \in [0.3, 2.7]$; with the assumption of spatially-varying coefficients, the local windows are $x \in [0.3 \cdot n, 0.3 \cdot (n+1)]$, ($n$=0,1,...,9), for each of which there are 400 spatial observation steps and 400 temporal observation steps in the domain $t \in [0.6, 5.4]$. $S_x$ and $S_t$ calculated are given in Table III, and it can be found that $S_t > S_x$ in all experiments, which means that the coefficients are temporally-varying. In addition, the best structure is obtained, which is shown in Table III.

For global meta-data, there are 300 temporal observation steps in the domain $t \in [0.5, 4.8]$ and 400 spatial observation steps in the domain $x \in [0.5, 2.5]$. The values of varying coefficients are calculated and DLGA-PDE (Coefficients) is performed. Results are shown in Table III. From the table, it is found that the general form of varying coefficients is discovered accurately.

**TABLE III.** Structure of parametric wave equation found using DLGA-PDE (Structure), and general parametric form found using DLGA-PDE (Coefficients), with clean and noisy data.

| Noise Level | $S_x$ | $S_t$ | Discovered Structure | Learned General Form by DLGA-PDE (Coefficients) | Error |
|---|---|---|---|---|---|
| Correct PDE | | | | $u_{tt} = (1+0.25\cos(1.5t))u_{xx}$ | |
| Clean Data | 0.4 | 0.8 | $u_{tt}, u_{xx}$ | $u_{tt} = (0.997+0.247\cos(1.494t))u_{xx}$ | $5.01 \times 10^{-6}$ |
| 1% Noise | 0.4 | 0.8 | $u_{tt}, u_{xx}$ | $u_{tt} = (0.999+0.244\cos(1.492t))u_{xx}$ | $2.36 \times 10^{-6}$ |
| 5% Noise | 0.4 | 0.7 | $u_{tt}, u_{xx}$ | $u_{tt} = (0.995+0.241\cos(1.499t))u_{xx}$ | $1.26 \times 10^{-5}$ |
| 10% Noise | 0.6 | 0.7 | $u_{tt}, u_{xx}$ | $u_{tt} = (0.995+0.245\cos(1.491t))u_{xx}$ | $1.35 \times 10^{-5}$ |
| 15% Noise | 0.8 | 0.9 | $u_{tt}, u_{xx}$ | $u_{tt} = (0.988+0.243\cos(1.498t))u_{xx}$ | $6.27 \times 10^{-5}$ |

**B. Discovery of parametric PDE with sparse data**

In this part, the performance of Adaptive DLGA-PDE for discovering parametric PDE under different data volumes is investigated. The Burgers equation, the convection-diffusion equation, and the wave equation are again considered. The settings of these PDEs are the same as those in Section III. Different amounts of data are randomly selected to train the neural network. For the Burgers equation and the convection-diffusion equation, 25,000 data, 15,000 data, 5,000 data, and 1,000 data are randomly chosen to form new datasets. For the wave equation, 30,000 data, 15,000 data, 5,000 data, and 1,000 data are randomly chosen. Adaptive DLGA-PDE is performed to discover these three PDEs, and results are shown in Table IV, V, and VI, respectively. From these tables, it can be seen that Adaptive DLGA-PDE is able to discover parametric PDEs in extremely sparse data (e.g., 1,000 data), which only accounts for nearly 2% of the total data. This means that Adaptive DLGA-PDE is robust to sparse data.

TABLE IV. Structure of parametric Burgers equation found using DLGA-PDE (Structure), and general parametric form found using DLGA-PDE (Coefficients), with different amounts of data.

| Volume of Data | $S_x$ | $S_t$ | Discovered Structure | Learned General Form by DLGA-PDE (Coefficients) | Error |
|---|---|---|---|---|---|
| Correct PDE | | | | $u_t = -uu_x + 0.02e^t u_{xx}$ | |
| 25,000 Data | 0.3 | 0.8 | $u_t, uu_x, u_{xx}$ | $u_t = -0.996uu_x + 0.01995e^{0.999t}u_{xx}$ | $1.15 \times 10^{-6}$ |
| 15,000 Data | 0.1 | 0.7 | $u_t, uu_x, u_{xx}$ | $u_t = -0.996uu_x + 0.01984e^{1.011t}u_{xx}$ | $1.29 \times 10^{-6}$ |
| 5,000 Data | 0.1 | 0.7 | $u_t, uu_x, u_{xx}$ | $u_t = -1.003uu_x + 0.02016e^{0.987t}u_{xx}$ | $7.77 \times 10^{-7}$ |
| 1,000 Data | 0.1 | 0.7 | $u_t, uu_x, u_{xx}$ | $u_t = -0.998uu_x + 0.01967e^{1.023t}u_{xx}$ | $1.74 \times 10^{-6}$ |

TABLE V. Structure of parametric convection-diffusion equation found using DLGA-PDE (Structure), and general parametric form found using DLGA-PDE (Coefficients), with different amounts of data.

| Volume of Data | $S_x$ | $S_t$ | Discovered Structure | Learned General Form by DLGA-PDE (Coefficients) | Error |
|---|---|---|---|---|---|
| Correct PDE | | | | $u_t = -(1 + 0.25\sin(x))u_x + u_{xx}$ | |
| 25,000 Data | 0.9 | 0.3 | $u_t, u_x, u_{xx}$ | $u_t = -(1.000 + 0.251\sin(1.001x))u_x + 0.999u_{xx}$ | $1.12 \times 10^{-6}$ |
| 15,000 Data | 0.9 | 0.2 | $u_t, u_x, u_{xx}$ | $u_t = -(1.000 + 0.252\sin(1.001x))u_x + 0.998u_{xx}$ | $3.21 \times 10^{-6}$ |
| 5,000 Data | 0.9 | 0.2 | $u_t, u_x, u_{xx}$ | $u_t = -(1.000 + 0.248\sin(0.999x))u_x + 0.996u_{xx}$ | $1.34 \times 10^{-5}$ |
| 1,000 Data | 0.9 | 0.4 | $u_t, u_x, u_{xx}$ | $u_t = -(1.000 + 0.250\sin(1.001x))u_x + 1.003u_{xx}$ | $1.74 \times 10^{-6}$ |

TABLE VI. Structure of parametric wave equation found using DLGA-PDE (Structure), and general parametric form found using DLGA-PDE (Coefficients), with different amounts of data.

| Volume of Data | $S_x$ | $S_t$ | Discovered Structure | Learned General Form by DLGA-PDE (Coefficients) | Error |
|---|---|---|---|---|---|
| Correct PDE | | | | $u_{tt} = (1+0.25\cos(1.5t))u_{xx}$ | |
| 30,000 Data | 0.4 | 0.8 | $u_{tt}, u_{xx}$ | $u_{tt} = (0.998+0.247\cos(1.493t))u_{xx}$ | $5.01\times10^{-6}$ |
| 15,000 Data | 0.5 | 0.6 | $u_{tt}, u_{xx}$ | $u_{tt} = (0.998+0.245\cos(1.497t))u_{xx}$ | $2.60\times10^{-6}$ |
| 5,000 Data | 0.5 | 0.8 | $u_{tt}, u_{xx}$ | $u_{tt} = (1.000+0.243\cos(1.485t))u_{xx}$ | $3.88\times10^{-6}$ |
| 1,000 data | 0.4 | 0.8 | $u_{tt}, u_{xx}$ | $u_{tt} = (0.999+0.239\cos(1.508t))u_{xx}$ | $3.50\times10^{-6}$ |

### C. Discovery of parametric PDE with an incomplete candidate library

Finally, the performance of Adaptive DLGA-PDE for discovering parametric PDEs with an incomplete candidate library is investigated. The ability of DLGA-PDE (Structure) to discover the structure of PDEs with an incomplete candidate library has been tested in our previous work, and additional details can be found in Xu et al. [18]. In this part, the ability of DLGA-PDE (Coefficients) to discover the general forms of coefficients in PDEs with incomplete basic genes is discussed.

Here, the Burgers equation is considered again. The basic gene of DLGA-PDE (Coefficients) is changed to be $[\alpha_i]\{1(0), k_1 t(1), \sin(k_2 t)(2), \cos(k_3 t)(3)\}$, with other conditions unchanged. In this case, the correct function $e^{k_4 t}$ is not contained in the basic gene, which means that it can only be produced by mutation. The best child in each generation is recorded, and the results are presented in Table VII. From the table, it can be seen that Adaptive DLGA-PDE discovered the incorrect form at first due to the absence of the correct function in the basic gene. However, after two generations, the correct function emerges via mutation, and Adaptive DLGA-PDE converges and discovers the correct form of the coefficient in PDE finally. This indicates that this method has the ability to discover parametric PDE with an incomplete candidate library. Additional details are provided in Appendix C, where the convection-diffusion equation and the wave equation are also tested, and satisfactory outcomes are obtained.

TABLE VII. Best child found in each generation when Adaptive DLGA-PDE is utilized to discover the parametric Burgers equation in the absence of a correct function in basic genes.

| Generations | Discovered Form of the PDE |
|---|---|
| 1 | $u_t = -0.996\, uu_x + 0.04447\sin(1.800t)u_{xx}$ |
| 2 | $u_t = -0.996\, uu_x + 0.04396\sin(1.848t)u_{xx}$ |
| 3 | $u_t = -0.996\, uu_x + 0.01792\, e^{1.167t} u_{xx}$ |
| 5 | $u_t = -0.996\,uu_x + 0.02076\,e^{0.936t}u_{xx}$ |
| 13 | $u_t = -0.996\,uu_x + 0.01923\,e^{1.057t}u_{xx}$ |
| 14 | $u_t = -0.996\,uu_x + 0.02004\,e^{0.992t}u_{xx}$ |
| 25 | $u_t = -0.996\,uu_x + 0.01998\,e^{0.997t}u_{xx}$ |
| 43 | $u_t = -0.996\,uu_x + 0.01995\,e^{0.999t}u_{xx}$ |
| 200 | $u_t = -0.996\,uu_x + 0.01995\,e^{0.999t}u_{xx}$ |

## IV SUMMARY AND OUTLOOK

In this work, we proposed a new framework combining adaptive methods and DLGA-PDE, called Adaptive DLGA-PDE, which aims to discover the general form of parametric PDEs from sparse noisy data. This algorithm utilizes DLGA-PDE (Structure) to discover the structure of PDEs, and employs DLGA-PDE (Coefficients) to identify the general form of varying coefficients. Faced with the three main challenges mentioned in Section I, Adaptive DLGA-PDE provides a systematic and comprehensive solution. In Adaptive DLGA-PDE, a neural network is utilized to calculate derivatives and generate meta-data, which solves the problem of sparse noisy data. Meanwhile, a genetic algorithm is applied to discover PDEs with an incomplete candidate library via mutation and combination of genes, which solves the problem of incomplete candidate library. Finally, a two-step adaptive method is developed to discover parametric PDEs, which solves the problem of varying coefficients. Numerical experiments demonstrated that Adaptive DLGA-PDE still performs well when the noise level is 15%, except for the KdV equation, which means that it is robust to noise and is able to handle various types of parametric PDEs.

Compared with PDEs with constant parameters, discovery of parametric PDEs is more complex and more easily affected by noise. Therefore, it is unprecedented in previous work that Adaptive DLGA-PDE is able to discover the correct form of parametric PDEs with high accuracy at 15% noise. The performance of our proposed method, however, is not satisfactory when discovering the parametric KdV equation. This is because the KdV equation has third-order derivatives, which may bring a relatively large error when calculating derivatives. To handle the problem of high-order derivatives, the weak form of PDEs mentioned in Section I may constitute a viable solution [12-13]. The weak form of PDEs is expressed by integral form, which will reduce the order of derivatives needed to be calculated. Integration is indispensable and important in the weak form, but it is difficult to calculate with discrete data. Therefore, our proposed algorithm is particularly suitable for performing integration because it can generate a large amount of meta-data on a regular grid, which assists in calculating integrals. Determination of how to combine the weak form and Adaptive DLGA-PDE constitutes a worthy topic of future work.

Numerical experiments also indicate that Adaptive DLGA-PDE is able to discover parametric PDEs with sparse data, even with only 2% of the total data. Moreover, the experiment in which the true term is not contained in the basic genes indicates that this method can discover parametric PDEs with an incomplete candidate library. Different from previous works, which are based on prior knowledge about whether the coefficients are spatially-varying or temporally-varying, Adaptive DLGA-PDE is able to identify the type of varying parameters, which greatly increases its potential for application.

To further investigate the stability of Adaptive DLGA-PDE and its application to address more complex problems, other experiments are carried out. Sensitivity analysis shows that the selection of local window size is important for discovering a correct structure. It is also observed that DLGA-PDE (Structure) is insensitive to the position of the local window with the correct assumption of varying coefficients. In total, Adaptive DLGA-PDE is very stable. Inspired by the stability of Adaptive DLGA-PDE, a more complex problem in which the structure of PDE is different in different domains is considered, and the results demonstrate that Adaptive DLGA-PDE performs well by producing local meta-data in different domains. Additional details about this problem are provided in Appendix D.

At the same time, Adaptive DLGA-PDE possesses certain limitations. For example, if the coefficient oscillates rapidly with a large amplitude, it may be difficult for DLGA-PDE (Structure) to discover the correct structure because coefficients with a large amplitude cannot be seen as a constant, even in a small local domain, which will lead to the failure to discover the true structure of PDE. In addition, the choice of hyper-parameters, including $\varepsilon$, $\varepsilon_1$ and $\varepsilon_2$ in the fitness function, is significant to discover the true PDE, but it is now decided by experience and is sophisticated to adjust. Meanwhile, if the form of varying coefficient is complex and cannot be represented by the combination of elementary functions, it may be difficult for Adaptive DLGA-PDE to discover the true general form of the varying coefficients. To solve this problem, discovering the form of Taylor expansion or an approximate substitution function may be a viable choice. Moreover, it may be difficult for Adaptive DLGA-PDE to directly discover PDEs whose coefficient is a random field. Dimension reduction techniques, such as Karhunen-Loeve expansion, singular value decomposition and auto-encoder, may be needed to parameterize the random field. Further investigation of these issues is necessary.

**Acknowledgements**

This work is partially funded by the National Natural Science Foundation of China (Grant No. 51520105005 and U1663208) and the National Science and Technology Major Project of China (Grant No. 2017ZX05009-005 and 2017ZX05049-003).

(2017).

[6] H. Chang and D. Zhang, Machine learning subsurface flow equations from data, Comput. Geosci. **23**, 895 (2019).

[7] M. Raissi, Deep hidden physics models: Deep learning of nonlinear partial differential equations, J. Mach. Learn. Res. **19**, 1 (2018).

[8] Z. Long, Y. Lu, and B. Dong, PDE-Net 2.0: Learning PDEs from data with a numeric-symbolic hybrid deep network, J. Comput. Phys. **399**, 108925 (2019).

[9] B. Xiong, H. Fu, F. Xu, and Y. Jin, Data-driven discovery of partial differential equations for multiple-physics electromagnetic problem, arXiv:1910.13531.

[10] T. Qin, K. Wu, and D. Xiu, Data driven governing equations approximation using deep neural networks, J. Comput. Phys. **395**, 620 (2019).

[11] K. Wu and D. Xiu, Data-driven deep learning of partial differential equations in modal space, arXiv:1910.06948.

[12] P. A. K. Reinbold, D. R. Gurevich, and R. O. Grigoriev, Data-driven discovery of partial differential equation models with latent variables, Phys. Rev. E **101**, 1 (2020).

[13] D. R. Gurevich, P. A. K. Reinbold, and R. O. Grigoriev, Robust and optimal sparse regression for nonlinear PDE models, Chaos **29**, (2019).

[14] M. Maslyaev, A. Hvatov, and A. Kalyuzhnaya, Data-driven PDE discovery with evolutionary approach, arXiv: 1903.08011.

[15] S. Atkinson, W. Subber, L. Wang, G. Khan, P. Hawi, and R. Ghanem, Data-driven discovery of free-form governing differential equations, arXiv:1910.05117.

[16] S. Rudy, A. Alla, S. L. Brunton, and J. N. Kutz, Data-driven identification of parametric partial differential equations, SIAM J. Appl. Dyn. Syst. **18**, 643 (2019).

[17] H. Xu, H. Chang, and D. Zhang, DL-PDE: Deep-learning based data-driven discovery of partial differential equations from discrete and noisy data, arXiv: 1908.04463.

[18] H. Xu, H. Chang, and D. Zhang, DLGA-PDE: Discovery of PDEs with incomplete candidate library via combination of deep learning and genetic algorithm, arXiv: 2001.07305.

[19] H. Chang and D. Zhang, Identification of physical processes via combined data-driven and data-assimilation methods, J. Comput. Phys. **393**, 337 (2019).

[20] Y. Chen and D. Zhang, Data assimilation for transient flow in geologic formations via ensemble Kalman filter, Adv. Water Resour. **29**, 1107 (2006).21

# APPENDIX A: DISCOVERY OF PARAMETRIC KDV EQUATION BY ADAPTIVE DLGA-PDE

In this section, the parametric KdV equation, which has a third-order derivative, is considered. Its form reads as follows:

$$u_t = -uu_x - b(x)u_{xxx},$$

$$b(x) = 0.0025(1+0.25\sin(\pi x))u_{xxx},$$

The KdV equation is solved numerically using the finite difference method, with the initial condition $u(0,x)=\cos(\pi x)$ and boundary condition $u(-1,t)=u(1,t)=0$, $t>0$. There are 201 temporal observation steps in the domain $t \in [0,1]$ and 512 spatial observation steps in the domain $x \in [-1,1)$, and thus the total number of data points is 102,912. 60,000 data are randomly chosen to train the neural network. The case with clean data is investigated. In this case, the type of varying parameters is assumed to be known. For local meta-data, there are 400 temporal observation steps in the local window $t \in [0.2, 0.8]$ and 400 spatial observation steps in the local window $x \in [-0.2, 0.2]$. For global meta-data, there are 300 temporal observation steps in the domain $t \in [0.05, 0.95]$ and 400 spatial observation steps in the domain $x \in [-0.95, 0.55]$. Adaptive DLGA-PDE is performed, and the results are shown in Table AI.

From the table, it can be seen that, although the structure and general form of varying coefficients are discovered successfully, the error is relatively large. There are many reasons contributing to the large error. Firstly, the parametric KdV equation has a third-order derivative, which is difficult to calculate accurately. Although automatic differentiation of the neural network is utilized to calculate derivatives, when faced with high-order derivatives, the error of calculated derivatives will still be large and affect the performance of Adaptive DLGA-PDE. Secondly, faced with the parametric KdV equation, which is a complex dynamic system, the structure of the neural network and the choice of activation function are important for learning its characteristics. The neural network $NN_1(x,t;\theta)$ in this work may not work well. Therefore, discovering a more appropriate structure of neural network and type of activation function will assist to learn the parametric KdV equation and calculate derivatives more accurately. Finally, for equations with high-order derivatives, more data are needed. Rudy et al. [16] investigated the Kuramoto-Sivashinsky (KS) equation, which has a fourth-order derivative. In their work, 262,144 data are used, and the results are robust to only 0.01% noise.

**TABLE AI.** Structure of KdV equation discovered using DLGA-PDE (Structure) and general parametric form found using DLGA-PDE (Coefficients).

|  | **Learned Structure** |  |
|---|---|---|
| **Correct PDE** | $u_t = -uu_x - 0.0025(1+0.25\sin(\pi x))u_{xxx}$ |  |
|  | **Learned Structure by DLGA-PDE (Structure)** | **Error** |
| **Discovered PDE** | $u_t = -1.013uu_x - 0.0024u_{xxx}$ | 0.657 |
|  | **Learned General Form by DLGA-PDE (Coefficients)** |  |

| | $u_t = -0.98uu_x - 0.0024(1+0.23\sin(2.96x))u_{xxx}$ | |

## APPENDIX B: SENSITIVITY ANALYSIS OF DLGA-PDE (STRUCTURE)

*1. Sensitivity to local window size*

To investigate the sensitivity of DLGA-PDE (Structure) to local window size, which is the size of the domain where local meta-data are generated, the Burgers equation is first taken as an example. In the sensitivity analysis, the type of varying parameter is supposed to be known. Local window size, which is referred as $L$, is changed. When generating local meta-data, 400 spatial observation points are uniformly selected from $t \in [0.5-\frac{L}{2}, 0.5+\frac{L}{2}]$. Other conditions are the same as those in Section IIIA. The structure discovered by DLGA-PDE (Structure) is recorded and compared with the true structure, which is shown in Fig. B1. In this figure, if the fitness of the discovered structure and the true structure are equal, this means that DLGA-PDE (Structure) has discovered the true structure. From this figure, it can be seen that the true structure can be discovered, even if the ratio of the window size to the entire length of the interval is nearly 50%.

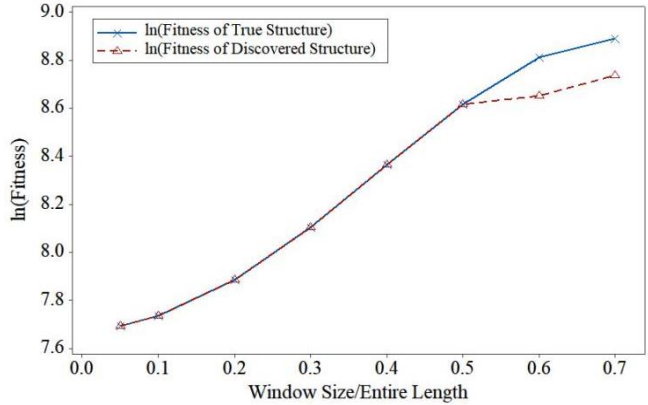

**FIG. B1.** Sensitivity to local window size in the parametric Burgers equation.

Next, the convection-diffusion equation is considered. When generating local meta-data, 400 spatial observation points are uniformly selected from $x \in [4-\frac{L}{2}, 4+\frac{L}{2}]$. Other conditions are the same as those in Section IIIB. The discovered structure is recorded and compared with the true structure, which is shown in Fig. B2. It is obvious that when the ratio of the window size to the entire length of the interval is less than 60%, the true structure can be successfully discovered.

Finally, the wave equation is considered. When generating local meta-data, 400 spatial observation points are uniformly selected from $t \in [3-\frac{L}{2}, 3+\frac{L}{2}]$. Other conditions are the same as those in Section IIIC. The discovered structure is recorded and compared with the true structure, which is shown in Fig. B3. It can be seen that the true structure can be discovered, even if the window size is as large as the entire length of the interval.

In general, DLGA-PDE (Structure) is able to discover the true structure when the local window size varies over a wide range, which indicates that it is not sensitive to the local window size.

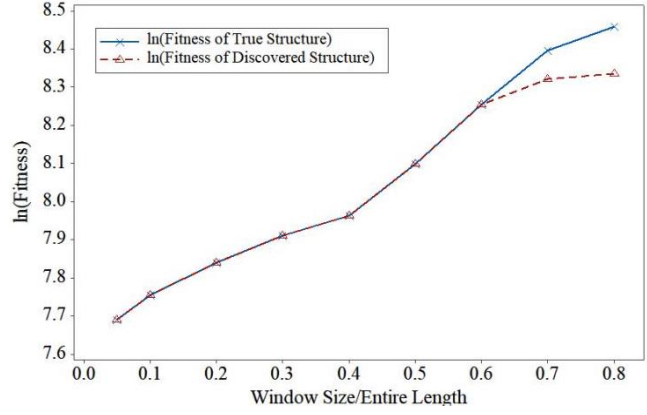

**FIG. B2.** Sensitivity to local window size in the parametric convection-diffusion equation.

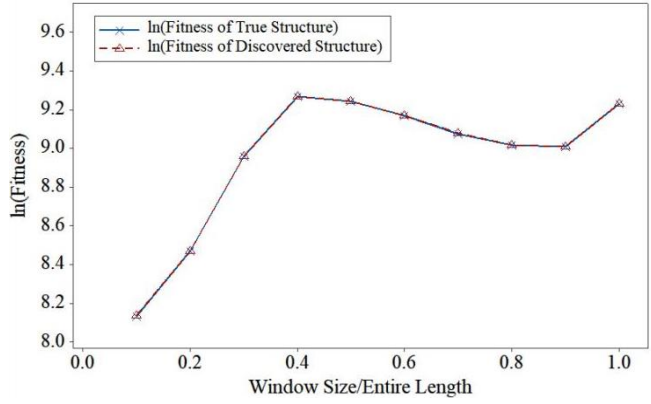

**FIG. B3.** Sensitivity to local window size in the parametric wave equation.

*2. Sensitivity to local window position*

Next, the sensitivity of DLGA-PDE (Structure) to local window position is investigated. The parametric Burgers equation with temporally-varying coefficients is first taken as an example. The basic case with clean data is investigated. Conditions are the same as those in Section IIIA. With the assumption of spatially-varying coefficients or temporally-varying coefficients, the structure discovered by DLGA-PDE (Structure) within each local window is recorded, respectively. The outcome is shown in Table BI. For this case, it can be calculated that $S_x$=0.3 and $S_t$=0.8. It is obvious that $S_t > S_x$. It can be found that, if the assumption is correct, structures discovered in multiple local windows are stable. Meanwhile, if the assumption is incorrect, the discovered structure changes according to the position of the local window. This means that DLGA-PDE (Structure) is insensitive to local window position with the correct assumption, but is sensitive to the local window position with the incorrect assumption.

**TABLE BI.** Sensitivity to local window position in the parametric Burgers equation.

| Assuming that Coefficients are Spatially-varying | Assuming that Coefficients are Temporally-varying |
|---|---|
| Correct Structure ||

| $u_t, uu_x, u_{xx}$ | | | |
|---|---|---|---|
| **Local Window in** $x$ | **Learned Structure** | **Local Window in** $t$ | **Learned Structure** |
| [0.0,0.1] | $u_t, uu_x$ | [0.0,0.1] | $u_t, uu_x, u_{xx}$ |
| [0.1,0.2] | $u_t, uu_x$ | [0.1,0.2] | $u_t, uu_x, u_{xx}, u_{xx}u_{xxx}$ |
| [0.2,0.3] | $u_t, uu_x$ | [0.2,0.3] | $u_t, uu_x, u_{xx}, u_{xx}u_{xxx}$ |
| [0.3,0.4] | $u_t, u, u^2 u_x$ | [0.3,0.4] | $u_t, uu_x, u_{xx}$ |
| [0.4,0.5] | $u_t, uu_x, u_{xx}, u^2 u_{xx}^2, u_x u_{xx}$ | [0.4,0.5] | $u_t, uu_x, u_{xx}$ |
| [0.5,0.6] | $u_t, uu_x, u_{xx}, u, u_{xx}u_{xxx}, u^2 u_{xxx}$ | [0.5,0.6] | $u_t, uu_x, u_{xx}$ |
| [0.6,0.7] | $u_t, uu_x, u_{xx}, uu_{xxx}$ | [0.6,0.7] | $u_t, uu_x, u_{xx}$ |
| [0.7,0.8] | $u_t, uu_x^2, u_{xx}$ | [0.7,0.8] | $u_t, uu_x, u_{xx}$ |
| [0.8,0.9] | $u_t, uu_x, u_{xx}$ | [0.8,0.9] | $u_t, uu_x, u_{xx}$ |
| [0.9,1.0] | $u_t, u$ | [0.9,1.0] | $u_t, uu_x, u_{xx}$ |

Next, the parametric convection-diffusion equation with spatially-varying coefficients is considered. Conditions are the same as those in Section IIIB. The basic case with clean data is investigated. With the assumption of spatially-varying coefficients or temporally-varying coefficients, the structure discovered by DLGA-PDE (Structure) within each local window is recorded, respectively. The outcome is shown in Table BII. For this case, it can be calculated that $S_x=0.9$ and $S_t=0.3$, which means $S_x>S_t$. It can be seen that DLGA-PDE (Structure) is more stable with the correct assumption.

**TABLE BII.** Sensitivity to local window position in the parametric convection-diffusion equation.

| Assuming that Coefficients are Spatially-varying | | Assuming that Coefficients are Temporally-varying | |
|---|---|---|---|
| **Correct Structure** | | | |
| $u_t, u_x, u_{xx}$ | | | |
| **Local Window in** $x$ | **Learned Structure** | **Local Window in** $t$ | **Learned Structure** |
| [0.0,0.8] | $u_t, u$ | [0.0,0.5] | $u_t, u_x, u_{xx}, uu_x$ |
| [0.8,1.6] | $u_t, u_x, u_{xx}$ | [0.5,1.0] | $u_t, u_x, u_{xx}, u_x^2, uu_{xxx}$ |
| [1.6,2.4] | $u_t, u_x, u_{xx}$ | [1.0,1.5] | $u_t, u_x, u_{xx}, u$ |
| [2.4,3.2] | $u_t, u_x, u_{xx}$ | [1.5,2.0] | $u_t, u_x, u_{xx}, u_x^2$ |
| [3.2,4.0] | $u_t, u_x, u_{xx}$ | [2.0,2.5] | $u_{tt}, u_{xx}^2$ |
| [4.0,4.8] | $u_t, u_x, u_{xx}$ | [2.5,3.0] | $u_{tt}, u_{xx}^2$ |
| [4.8,5.6] | $u_t, u_x, u_{xx}$ | [3.0,3.5] | $u_{tt}, u_{xx}^2$ |
| [5.6,6.4] | $u_t, u_x, u_{xx}$ | [3.5,4.0] | $u_{tt}, uu_x$ |
| [6.4,7.2] | $u_t, u_x, u_{xx}$ | [4.0,4.5] | $u_{tt}, uu_x$ |
| [7.2,8.0] | $u_t, u_x, u_{xx}$ | [4.5,5.0] | $u_{tt}, u_{xx}$ |

Finally, the parametric wave equation with temporally-varying coefficients is considered. Conditions are the same as those in Section IIIC with clean data. With the assumption of spatially-varying coefficients or temporally-varying coefficients, the structure discovered by DLGA-PDE (Structure) within each local window is recorded, respectively. The outcome is shown in Table BIII. For this case, it can be calculated that $S_x$=0.4 and $S_t$=0.8, which means $S_x<S_t$. It can be seen that DLGA-PDE (Structure) is more stable with the correct assumption.

In summary, DLGA-PDE (Structure) is insensitive to local window position with the correct assumption of the varying parameter, but is sensitive to local window position with the incorrect assumption.

**TABLE BIII.** Sensitivity to local window position in the parametric wave equation.

| Assuming that Coefficients are Spatially-varying | | Assuming that Coefficients are Temporally-varying | |
|---|---|---|---|
| Correct Structure | | | |
| $u_{tt}, u_{xx}$ | | | |
| **Local Window in *x*** | **Learned Structure** | **Local Window in *t*** | **Learned Structure** |
| [0.0,0.3] | $u_t, u$ | [0.0,0.6] | $u_t, u_{xx}, u_{xx}u_{xxx}, u_{xx}^2$ |
| [0.3,0.6] | $u_{tt}, uu_{xx}^2 u_{xxx}^2, u_{xx}$ | [0.6,1.2] | $u_{tt}, u_{xx}$ |
| [0.6,0.9] | $u_{tt}, u_{xx}^3 u_{xxx}^2, u_{xx}$ | [1.2,1.8] | $u_t, u_{xx}, u_{xx}u_{xxx}, u_{xx}^2$ |
| [0.9,1.2] | $u_{tt}, u_{xx}$ | [1.8,2.4] | $u_{tt}, u_{xx}$ |
| [1.2,1.5] | $u_{tt}, u_{xx}$ | [2.4,3.0] | $u_{tt}, u_{xx}$ |

| [1.5,1.8] | $u_{tt}, u_{xx}$ | [3.0,3.6] | $u_{tt}, u_{xx}$ |
| --- | --- | --- | --- |
| [1.8,2.1] | $u_{tt}, u_{xx}$ | [3.6,4.2] | $u_{tt}, u_{xx}$ |
| [2.1,2.4] | $u_{tt}, u_{xx}u_{xxx}^2, u_{xx}$ | [4.2,4.8] | $u_{tt}, u_{xx}$ |
| [2.4,2.7] | $u_{tt}, u_{xx}^3 u_{xxx}^2, u_{xx}$ | [4.8,5.4] | $u_{tt}, u_{xx}$ |
| [2.7,3.0] | $u_t, u_x u_{xx} u_{xxx}$ | [5.4,6.0] | $u_{tt}, u_{xx}$ |

## APPENDIX C: DISCOVERY OF PARAMETRIC PDES WITH AN INCOMPLETE CANDIDATE LIBRARY

To further investigate the performance of Adaptive DLGA-PDE when discovering parametric PDEs with an incomplete candidate library, the convection-diffusion equation and the wave equation are also considered.

For the convection-diffusion equation, the basic gene of DLGA-PDE (Coefficients) is changed to be $[a_i]\{1(0), k_1x(1), \cos(k_3x)(3), e^{k_4x}(4)\}$, with other conditions unchanged. Here, the correct function $\sin(k_2x)$ is not contained in the basic gene, which means that it can only be produced by mutation. The best child in each generation is recorded, and the results are shown in Table CI.

**TABLE CI.** Best child found in each generation when Adaptive DLGA-PDE is utilized to discover the parametric convection-diffusion equation in the absence of a correct term in basic genes.

| Generations | Discovered Form of the PDE |
| --- | --- |
| 1 | $u_t = -1.020 u_x + 0.999 u_{xx}$ |
| 2 | $u_t = -1.089 \cos(0.081x) u_x + 0.999 u_{xx}$ |
| 3 | $u_t = -(1.011 + 0.247 \sin(0.953x)) u_x + 0.999 u_{xx}$ |
| 5 | $u_t = -(1.010 + 0.248 \sin(0.958x)) u_x + 0.999 u_{xx}$ |
| 7 | $u_t = -(0.994 + 0.247 \sin(1.030x)) u_x + 0.999 u_{xx}$ |
| 14 | $u_t = -(0.999 + 0.251 \sin(1.007x)) u_x + 0.999 u_{xx}$ |
| 24 | $u_t = -(1.000 + 0.251 \sin(1.001x)) u_x + 0.999 u_{xx}$ |
| 200 | $u_t = -(1.000 + 0.251 \sin(1.001x)) u_x + 0.999 u_{xx}$ |

For the parametric wave equation, the basic gene of DLGA-PDE (Coefficients) is changed to be $[a_i]\{1(0), k_1t(1), \sin(k_2t)(3), e^{k_4t}(4)\}$, with other conditions unchanged. Here, the correct function $\cos(k_3t)$ is not contained in the basic gene. The best child in each generation is recorded, and the results are presented in Table CII.

**TABLE CII.** Best child found in each generation when Adaptive DLGA-PDE is utilized to discover the parametric wave equation in the absence of a correct term in basic genes.

| Generations | Discovered Form of the PDE |
|---|---|
| 1 | $u_{tt} = (6.284te^{-2.18t} + 0.294t)u_{xx}$ |
| 2 | $u_{tt} = (1.683te^{-0.89t} + 0.274t)u_{xx}$ |
| 3 | $u_{tt} = (1.003 + 0.245\cos(1.444t))u_{xx}$ |
| 4 | $u_{tt} = (1.000 + 0.247\cos(1.469t))u_{xx}$ |
| 7 | $u_{tt} = (0.997 + 0.248\cos(1.502t))u_{xx}$ |
| 105 | $u_{tt} = (0.997 + 0.247\cos(1.492t))u_{xx}$ |
| 116 | $u_{tt} = (0.997 + 0.247\cos(1.495t))u_{xx}$ |
| 135 | $u_{tt} = (0.997 + 0.247\cos(1.494t))u_{xx}$ |
| 200 | $u_{tt} = (0.997 + 0.247\cos(1.494t))u_{xx}$ |

APPENDIX D: THE APPLICATION OF ADAPTIVE DLGA-PDE IN PDES WITH DIFFERENT STRUCTURES IN DIFFERENT DOMAINS

Inspired by the stability of Adaptive DLGA-PDE, a more complex problem in which the structure of PDE is different in different domains is considered, which is described as:

$$u_t = a(x)u_x + u_{xx}$$

$$a(x) = \begin{cases} -(1 + \frac{\sin(x)}{4}), & 0 \le x < 2\pi \\ 0, & 2\pi \le x < 8 \end{cases}$$

To obtain training data, this PDE is solved numerically using the precise integration method with the initial condition $(8-x)\cdot\sin(x)$ and boundary condition $u(0,t)=u(8,t)=0$, $t>0$. There are 250 temporal observation steps in the domain $t \in [0,5)$ and 201 spatial observation steps in the domain $x \in [0,8]$, and thus the total number of data points is 50,250, of which 30,000 data are randomly chosen to train the neural network. Here, the type of varying parameter is assumed to be known.

To discover the structure in different domains, a mobile local window is adopted. To generate local meta-data, the length of the local window is fixed to be 0.5. The middle point of the window is referred to as *M*. There are 400 spatial observation steps in the domain $x \in [M-0.25, M+0.25]$ and 400 temporal observation steps in the domain $t \in [0.5, 4.5]$. The local window moves as *M* increases, and the structures discovered by DLGA-PDE (Structure) on different domains are recorded, which is shown in Table DI. From the table, it is obvious that the entire interval is divided into two sections, one is $x \in [0,6]$ and the other is $x \in [6.5,8]$. This means that the translation point is in the interval $x \in [6,6.5]$.

**TABLE DI.** Discovered structure by DLGA-PDE (Structure) on different local window positions.

| Local Window | Discovered Structure |
|---|---|
| $x \in [1,1.5]$ | $u_t, u_x, u_{xx}$ |
| $x \in [2.5,3]$ | $u_t, u_x, u_{xx}$ |
| $x \in [4,4.5]$ | $u_t, u_x, u_{xx}$ |
| $x \in [5,5.5]$ | $u_t, u_x, u_{xx}$ |
| $x \in [5.5,6]$ | $u_t, u_x, u_{xx}$ |
| $x \in [6,6.5]$ | $u_t, u, u_x, u_{xxx}$ |
| $x \in [6.5,7]$ | $u_t, u_{xx}$ |
| $x \in [7,7.5]$ | $u_t, u_{xx}$ |

After determining these two sections of domains, global meta-data are produced in respective domains. For $x \in [0,6]$, there are 400 spatial observation steps in the domain $x \in [0,5.5]$. For $x \in [6.5,8]$, there are 400 spatial observation steps in the domain $x \in [6.5,7.5]$. For both situations, there are 300 temporal observation steps in the domain $t \in [0.2,4.8]$. The values of varying coefficients are calculated, DLGA-PDE (Coefficients) is performed to identify the general form, and the results are presented in Table DII. It can be seen that Adaptive DLGA-PDE discovered the PDEs in different domains with high accuracy.

**TABLE DII.** Discovered PDE by Adaptive DLGA-PDE in different domains.

| Domain of Meta-data | | Learned Equation |
|---|---|---|
| $x \in [0.5,5.5]$ | Correct PDE | $u_t = -(1+0.25\sin(x))u_x + u_{xx}$ |

| | Learned Equation | $u_t = -(0.998 + 0.250\sin(x))u_x + 1.000 u_{xx}$ |
|---|---|---|
| $x \in [6.5, 7.5]$ | Correct PDE | $u_t = u_{xx}$ |
| | Learned Equation | $u_t = 0.989 u_{xx}$ |